\documentclass[a4paper,12pt,oneside]{amsart}
\usepackage{authblk}

\usepackage[textheight=25 cm, textwidth=17cm, headheight=50pt, headsep=10pt, footskip=32pt]{geometry}
\usepackage{algorithm}
\usepackage{algpseudocode}
\usepackage{geometry}
\usepackage{epstopdf}
\usepackage{mathtools}
\usepackage[cmtip,all]{xy}
\usepackage{setspace} 

\usepackage{graphicx}
\usepackage{amssymb}
\usepackage{amsfonts}
\usepackage{amsmath}
\usepackage{mathtools}
\usepackage{epstopdf}
\usepackage{color}
\usepackage[all,cmtip]{xy}
\usepackage{array}
\usepackage{tikz-cd}
\usepackage{tabularx}
\usepackage{fancyhdr}
\usepackage{booktabs}
\usepackage[normalem]{ulem}

\usepackage{caption}
\usepackage{subcaption}

\usepackage{lipsum}

\usepackage{bm,dsfont}
\usepackage{multirow}
\usepackage{enumitem}
\usepackage{color}
\definecolor{red}{rgb}{0.70,0.13,0.13}
\definecolor{green}{rgb}{0.13,0.55,0.13}
\definecolor{blue}{rgb}{0.25, 0.41, 0.88}
\usepackage{hyperref}
\hypersetup{
colorlinks = true,
linkcolor = blue,
citecolor = green,
urlcolor = blue}

\newcommand{\dsE}{\mathbb{E}}

\newcommand{\scF}{\mathcal{F}}

\newcommand{\scN}{\mathcal{N}}

\newcommand{\vect}[1]{{\bm{#1}}}

\usepackage{times} 

\usepackage[english]{babel}
\usepackage[numbers,sort&compress]{natbib}

\theoremstyle{plain}

\newtheorem{thm}{Theorem}[section]

\newtheorem{defi}[thm]{Definition}

\newtheorem*{prop*}{Proposition}

\newtheorem{prop-defi}[thm]{Proposition-Definition}
\newtheorem{thm-defi}[thm]{Theorem-Definition}
\newtheorem{lem-defi}[thm]{lema-Definition}

\newenvironment{itemize*}%
  {\vspace{-5pt}\begin{itemize}%
    \setlength{\itemsep}{0pt}%
    \setlength{\parskip}{0pt}}%
  {\end{itemize}\vspace{-5pt}}
 \newenvironment{enumerate*}%
  {\vspace{-5pt}\begin{enumerate}[label={(\arabic*)}]%
    \setlength{\itemsep}{0pt}%
    \setlength{\parskip}{0pt}}%
  {\end{enumerate}\vspace{-5pt}}
\usepackage{tikz-cd}

\begin{document}

\title{Renormalization Group flow, Optimal Transport and Diffusion-based Generative Model}
\maketitle

\begin{center}
\author{}{Artan Sheshmani${}^{1,2,3}$, Yi-Zhuang You${}^{4}$, Baturalp Buyukates${}^{5}$, Amir Ziashahabi${}^{5}$ \\and Salman Avestimehr${}^{5}$}
\end{center}
\address{${}^1$  MIT, Institute for Artificial Intelligence and Fundamental Interactions, 77 Mass Ave, Cambridge, MA 02138}
\address{${}^2$ Beijing Institute of Mathematical Sciences and Applications, No. 544, Hefangkou Village, Huaibei Town, Huairou District, Beijing 101408}
\address{${}^3$  Harvard University, Jefferson Laboratory, 17 Oxford St, Cambridge, MA 02138}
\address{${}^4$ University of California San Diego, Department of Physics, Condensed Matter Group, UC San Diego 9500 Gilman Dr. La Jolla, CA 92093}
\address{${}^5$ University of Southern California, Department of Electrical and Computer Engineering, Los Angeles, CA 90089}
\date{\today}

\begin{abstract}
Diffusion-based generative models represent a forefront direction in generative AI research today. Recent studies in physics have suggested that the renormalization group (RG) can be conceptualized as a diffusion process. This insight motivates us to develop a novel diffusion-based generative model by reversing the momentum-space RG flow. We establish a framework that interprets RG flow as optimal transport gradient flow, which minimizes a functional analogous to the Kullback-Leibler divergence, thereby bridging statistical physics and information theory. Our model applies forward and reverse diffusion processes in Fourier space, exploiting the sparse representation of natural images in this domain to efficiently separate signal from noise and manage image features across scales. By introducing a scale-dependent noise schedule informed by a dispersion relation, the model optimizes denoising performance and image generation in Fourier space, taking advantage of the distinct separation of macro and microscale features. Experimental validations on standard datasets demonstrate the model's capability to generate high-quality images while significantly reducing training time compared to existing image-domain diffusion models. This approach not only enhances our understanding of the generative processes in images but also opens new pathways for research in generative AI, leveraging the convergence of theoretical physics, optimal transport, and machine learning principles.
\smallskip



\end{abstract}
\tableofcontents

\section*{Introduction}

Deep generative models have been applied to various domains, including image generation, text generation, and audio synthesis. They have been used for tasks such as image generation, data augmentation, anomaly detection, and even as tools for understanding the underlying structure of the data. The diffusion generative models are a special type of generative model aiming to learn the underlying distribution of the data by iteratively applying a diffusion process to a simple noise distribution. They model the data as the result of a sequence of transformations that gradually corrupt the noise until it resembles the desired data distribution. The core idea behind the diffusion process is to introduce noise into the generated samples and then gradually reduce the noise level over multifold steps. To summarize it: each step of the diffusion process consists of two operations: an update step and a sampling step. During the update step, the model takes a noisy sample and updates it to reduce the noise level. This is typically done using a neural network that takes the current noisy sample and produces an estimate of the noise that needs to be subtracted. During the sampling step, the model samples from the updated noisy sample to generate a new sample. This step introduces stochasticity into the process and allows for the exploration of the data distribution. The diffusion process is performed for multiple steps, progressively reducing the noise level until the generated sample resembles the target data distribution. The model is trained by optimizing the likelihood of the data given the diffusion process.

One advantage of diffusion generative models is that they allow one to generate high-quality samples and capture complex data distributions. They have shown impressive results in various domains, including image generation, text generation, and audio synthesis. The training process of diffusion models can be computationally intensive, however recent advancements have introduced techniques to improve training efficiency, such as using reversible networks and applying diffusion models in a hierarchical manner. The current article is an attempt to create a diffusion generative model, based on ideas borrowed from mathematics (algebraic geometry, category theory, and theory of optimal transports) and physics (theory of renormalization group flows). A part of this endeavor has already been established by the work of the first two named authors, that is, the construction of a flow-based auto-encoding and auto-decoding algorithm based on category theory and renormalization group flows in quantum field theory. 

Several recent works \cite{2023MLS&T...4d5011B, 2022arXiv221211379B, 2022arXiv220412939B} have pointed out that the renormalization group (RG) can be conceptualized as an optimal transport of complicated probability distributions towards trivial distributions, described as a diffusion process. Following the early idea that inverse RG can be viewed as a generative model \cite{2013arXiv1301.3124B, 2014arXiv1410.3831M, PhysRevResearch.2.023369}, there has been proposals of using generative model to learn the optimal RG flow \cite{2015NJPh...17h3005B, 2023arXiv230611054H, 2023arXiv230812355C}. In this work, we will explore the opposite direction of developing diffusion-based generative models by reversing the RG flow. 

In \cite{RG} the authors constructed an algorithmic architecture capable of rapid classification of large-size data with few labels, as well as the discovery of new data that could potentially have similar characteristics to labeled data. The authors further showed as an example, that the construction can have applications in the field of ``\textit{Sequence-to-Function-Mapping}" which is currently a highly attractive focus of development in the Bio-Tech industry. The construction of the ``RG-flow based categorifier" borrows ideas from the theory of renormalization group flows (RG) in quantum field theory, holographic duality, and hyperbolic geometry, and mixing them with neural ODE's (ordinary differential equations). The authors constructed a new algorithmic natural language processing (NLP) architecture, called the RG-flow categorifier or for short the RG categorifier, which is capable of data classification and generation in all layers. The authors in \cite{RG} showed that the RG categorifier is capable of 
\begin{itemize} 
\item Combining representation and generative learning of data sets
\item Unsupervised classification of data sets 
\item Discovering fine and coarse features within the dataset
\item Precise analysis of data sets with their given functional characteristics, in particular, it enjoys trackable likelihood estimation capability
\item Being computationally efficient
\item Having explainability features, that is, capable of exploring explicitly how local structures in the data sets, induce their global features 
\end{itemize}

Despite having the advantages mentioned above, the RG categorifier shows low efficiency in speed which led the authors of the current article to envision the construction of a similar, yet less computationally intensive auto-encoder and decoder which works based on Diffusion dynamics in optimal transport theory. 

Optimal transport theory, also known as transportation theory or Monge-Kantorovich theory, is a mathematical framework for studying the transportation of mass from one location to another while minimizing the cost of transportation. This theory has applications in a variety of fields, including economics, physics, and image processing.

At its core, optimal transport theory seeks to find the optimal way to move a certain amount of mass from one location to another, subject to certain constraints such as the cost of transportation or the total amount of mass that can be transported. This is done by formulating the problem as an optimization problem and solving it using various mathematical techniques such as linear programming, convex analysis, and partial differential equations.

One of the key concepts in optimal transport theory is the Wasserstein distance, which measures the distance between two probability distributions by computing the minimum amount of work required to transform one distribution into the other. This concept has important applications in image processing, where it can be used to compare two images and measure the amount of deformation required to transform one image into the other.

When it comes to artificial intelligence (AI) and machine learning (ML), one of the main areas of application of optimal transport theory is in generative models, where optimal transport can be used to measure the distance between probability distributions and help guide the training of models.

In particular, optimal transport theory has been used to develop a new class of generative models called Wasserstein generative adversarial networks (WGANs), which improve upon traditional GANs by using the Wasserstein distance to measure the distance between the generated and real data distributions. This results in more stable training and better sample quality.

Optimal transport theory has also been used in image synthesis, where it can be used to learn the mapping between two images or image domains. For example, optimal transport theory has been used to develop domain adaptation methods, where a model is trained on one set of images and then transferred to a different set of images by learning the optimal transport plan between the two image domains. Another area of application is in natural language processing (NLP), where optimal transport theory can be used to measure the similarity between two text documents or the distance between two-word embeddings. This has led to the development of new techniques for text classification and clustering based on optimal transport. 

We apply the theory of optimal transport to a Fourier space diffusion problem and propose the Frequency Domain Diffusion Model (FDDM) as an alternative to image domain diffusion models. FDDM leverages the natural separation of components in the frequency domain and utilizes a scale-dependent noise schedule to intelligently add/remove noise during the diffusion process for efficient image generation. Combined with a JPEG-inspired design, FDDM achieves a computational speedup of $2.7-8.5\times$, with a modest impact on image quality.

This article is organized as follows: 
In Section \ref{Mong-Kont} we review the Monge-Kantorovich formulation in optimal transport theory. Then in Section \ref{RG-MK} we show that there is a close and deep relationship between RG flow theory and Monge-Kantorovich formulation, that is replacing the probabilistic metric used to optimize the Wasserstein distance function, with a certain derivative of the correlation matrix in RG dynamics, turns the RG flow equation into a Benamou-Brenier discretized diffusion equation on space of probability distributions on a given fixed space. After this mathematical construction, we apply the theory of optimal transport to Fourier space associated with images in Section \ref{F-Image}, and show the performance of an auto encoding and decoding algorithm which implements Benamou-Brenier discretized diffusion to encode and decode data.  

\section{Background: Optimal transport theory, Monge and Kantorovich formulations}\label{Mong-Kont}
In this section, we will first review the formulation of renormalization group (RG) flow as optimal transport developed by Cotler and Rezchikov \cite{Cotler2023R2202.11737}. In this formulation, the RG flow can be viewed as a trivializing flow of a probability distribution, described by a diffusion equation. The inverse diffusion can then be used to design a diffusion-based generative model.

 Let $X$ and $Y$ be separable metric spaces with positive measure $\mu_{X}$ and $\mu_{Y}$ respectively. The Monge formulation of optimal transportation problem is to find a ``\textit{transport}" map $T: X \to Y$  such that for any measure subset $S\subset X$ we have that $$\displaystyle{\int}_{S\subset X}\mu_{X}dx=\displaystyle{\int}_{T(S)}\mu_{Y}dy,$$that is, using the pullback pushforward formula, we have that $$T_{*}\mu_{X}=\mu_{Y}.$$The map $T$ will be called an optimal transport if attains the infimum value of the expected cost function associate to the transport, that is$$\text{inf} \{\displaystyle{\int}_{X\times Y}c(x,y)d \eta_{X\times Y}\mid \eta_{X\times Y}\in \Gamma(\mu_{X}, \mu_{Y}) \},$$where $c(x,y): X\times Y\to [0,\infty]$ is a Borel measurable function and $\Gamma(\mu_{X}, \mu_{Y})$ denotes the set of probability measures on $X\times Y$ with marginal probability measures $\mu_{X}$ on $X$ and $\mu_{Y}$ on $Y$. As we elaborated the pushforward of Borel measure $\mu_{X}$ under transport map $T$ is the Borel measure $\mu_{Y}$. When the map $T$ is given by a $C^1$-smooth map, this identity is realized in appropriate local coordinate charts as 
 \begin{equation}\label{nonlinear}
 \mu_{X}(x)=\mu_{Y}(T(x))\mid \text{det}(\partial_{x} T(x))\mid
 \end{equation}
This nonlinear condition can be relaxed to a KL divergence regularization.
\begin{equation}
\mathcal{L}(T, \lambda)=\displaystyle{\int}_{X}c(x,T(x))+\lambda \log\frac{\mu_{X}(x)}{\mu_{Y}(T(x))\mid \text{det}(\partial_{x} T(x))\mid}
\end{equation} 
\subsection{Generalization to Fuzzy Transport}
The expected cost function above relies on the assumption that the transport map $T: X\to Y$ is given deterministically. A fruitful generalization of this construction is to assume that the morphism $T$ is given as a conditional probability condition $\mu_{Y}(y\mid x)$.  The conditional distribution satisfies the identity $$\displaystyle{\int}_{Y}\mu_{X}(x)\mu_{Y}(y\mid x)dx.$$ On the other hand the conditional distribution $\mu_{Y}(y)$ is related to the joint distribution $\pi(x,y)$ as $\mu_{Y}(y\mid x)=\frac{\pi(x,y)}{\mu_X(x)}.$Therefore we obtain$$\mu_{Y}(y)=\displaystyle{\int}_{X} \pi(x,y) \,\,\,dx$$and similarly $$\mu_{X}(x)=\displaystyle{\int}_{Y} \pi(x,y) \,\,\,dy$$ 
These imply that
\begin{align}
&\mu_{Y}(y\mid x)=\frac{\pi(x,y)}{\mu_{X}(x)}=\frac{\pi(x,y)}{\displaystyle{\int}_{Y} \pi(x,y) \,\,\,dy}\notag\\
&\mu_{X}(x\mid y)=\frac{\pi(x,y)}{\mu_{Y}(y)}=\frac{\pi(x,y)}{\displaystyle{\int}_{X} \pi(x,y) \,\,\,dx}\notag
\end{align}

\subsection{Kantorovich Formulation}
Given two spaces $X$ and $Y$ with positive measures $\mu_{X}$ on $X$ and	$\mu_{Y}$ on $Y$, the Kantorovich formulation of the optimal transport asks to ﬁnd a positive measure $\pi_{X,Y}$ on $X \times Y$, such that the following conditions are satisfied:\\
1. The pushforward of $\pi_{X,Y}$ to $X$ is $\mu_{X}$ and the pushforward of $\pi_{X,Y}$ to $Y$ is $\mu_{Y}$. As the pushforward of the constructible measurable function $\pi_{X,Y}$  is the integral along the fibers of the projective morpshism $\pi_{1}: X\times Y\to X$ and $\pi_{2}: X\times Y\to Y$ respectively, one can say that the compatibility of measures induces the marginal probability conditions on $X$ and $Y$ respectively $$\mu_{Y}(y)=\displaystyle{\int}_{X} \pi(x,y) \,\,\,dx$$and similarly $$\mu_{X}(x)=\displaystyle{\int}_{Y} \pi(x,y) \,\,\,dy.$$The measure $\pi_{X}$ minimizes the expected cost
\begin{equation}\label{kanto}
\mathcal{L}(\pi_{X,Y})=\displaystyle{\int}_{X\times Y}c(x,y)\pi_{X,Y}(x,y)\,\,\,dx\, dy
\end{equation} 
for some cost function $c: X\times Y\to [0,\infty]$. This is a convex optimization problem, known as the primal Kantorovich problem, which is easier to solve compared to the Monge formulation.
\subsection{Kantorovich Duality}
The primal Kantorovich problem \eqref{kanto} admits a dual formulation, known as the dual Kantorovich problem,
\begin{align}
&\text{max}_{}\mathcal{L}(\pi_{X,Y})[\phi, \psi]=\int_{X} \mu_{X}(x) \phi(x)+\int_{Y} \mu_{Y}(y) \psi(y)dy\notag\\
&
\phi(x)+\psi(y)\leq c(x,y)
\end{align}
\subsection{Wasserstein Distance}
\begin{defi}
Given a metric space $(X,g)$ the optimum value of the Kantorovich problem defines the Wasserstein distance
\begin{equation}
\mathcal{W}(\mu_{X})=\text{inf}_{\pi\in \Gamma(\mu(x), \mu(y))} \left( \displaystyle{\int}_{X\times X}\pi(x,y)d_{g}(x,y)dx\,\,dy\right)^{1/2}
\end{equation} 
where $d_{g}(x,y)$ is the $g$-distance between any two points $x,y\in X$, and $\Gamma(\mu(x), \mu(y))$ is the space of probability distributions $\pi$ on $X\times X$ such that $$\mu(x)=\displaystyle{\int}_{X} \pi(x,y) \,\,\,dy$$and similarly $$\mu(y)=\displaystyle{\int}_{X} \pi(x,y) \,\,\,dx.$$
\end{defi}
\begin{defi}
Given a scalar density functional $F: dens(X)\to \mathbb{R}$, its Wasserstein gradient $\delta F[\mu]/ \delta \mu$ is defined by 
\begin{equation}
F[\mu+\delta \mu]=F[\mu]+\left<\frac{\delta F[\mu]}{\delta \mu}, \delta \mu\right>_{\mathcal{W}}+O(\delta \mu^2)
\end{equation}
where $\left<\frac{\delta F[\mu]}{\delta \mu}, \delta \mu\right>_{\mathcal{W}}$ is the Wasserstein distance evaluated on the two infinitesimal vectors $\frac{\delta F[\mu]}{\delta \mu}, \delta \mu$ defined as sections of the tangent sheaf over $dens(X)$, the space of probability distributions on $X$.
\end{defi}
Given this definition one can formulate Wasserstein gradient of energy functionals over the space of probability distributions on $X$. Our aim here is to show that Wasserstein variation of such integrals induces the diffusion dynamics on space of probability distributions on $X$. That is to say that the stochastic diffusion equation can be realized as the gradient flow on Wasserstein space. First we define the energy functional, then the entropy and interactions functionals. The diffusion equations are obtained by computing the Wasserstein gradient of the sum of energy, entropy and interaction functionals.
\begin{defi}\label{energy}
Define the energy functional 
\begin{equation}
E[\mu]=\int_{X}\mu(x)\epsilon(x)dx
\end{equation}
where $\epsilon(x)$ denotes the energy of $x$. Its Wasserstein gradient is given by
\begin{equation}
\frac{\delta E[\mu]}{\delta \mu(x)}= -\partial_{x}\cdot g^{-1}\cdot(\mu(x)\partial_{x}\epsilon(x))
\end{equation}
\end{defi}
Now we define the entropy functional
\begin{defi}\label{entropy}
Define the entropy functional 
\begin{equation}
S[\mu]=-\int_{X} \mu(x)\log(\mu(x))dx
\end{equation}
\end{defi}
\begin{defi}\label{free}
Consider Definition \ref{energy} and Definition \ref{entropy}. The free energy functional is defined as 
\begin{equation}
F[\mu]=E[\mu]-TS[\mu]=\int_{X} \left(\mu(x)\epsilon(x)+ T \log(\mu(x)) \right)dx
\end{equation}
\end{defi}
Now we are ready to derive a diffusion equation as the gradient flow of the Free energy functional defined above. 
\begin{defi}
Consider Definition \ref{free}. The Fokker-Plank diffusion relaxation equation is defined as the Wasserstein gradient flow applied to the free energy functional. That is
\begin{equation}\label{Fokker}
\frac{\delta F[\mu]}{\delta \mu(x,t)}= -T \partial_{x}\cdot g^{-1}\cdot \partial_{x}\mu(x, t)+\partial_{x}\cdot g^{-1}\cdot (\mu(x,t)\partial_{x}\epsilon(x))
\end{equation}
where the first term describes the diffusion, and the second term describes the relaxation.
\end{defi}
\section{Benamou-Brenier formulation of optimal transports and discretized diffusion equation}\label{B-B}
On a manifold $X$, define the Wasserstein distance between probability  measures $\mu, \nu$ as 
\begin{equation}
W_{2}^{2}=\text{inf}_{\mu_{t}, \nu_{t}}\{\displaystyle{\int}_{0}^{1}\mid\mid \nu_{t}\mid\mid_{L^{2}(\mu_{t})}dt\}
\end{equation}
where the infimum runsa over all 2-absolutely continuous curves $\mu_{t}, t\in [0,1]$ in the $L^{2}$-Wasserstein space, which satisfy the continuity equation 
\begin{equation}
\frac{d\mu_{t}}{dt}+\nabla \cdot (\nu_{t}\mu_{t})=0.
\end{equation}
Here $\mu_{t}$ satisfies the boundary condition $\mu_{0}=\mu$ and $\mu_{1}=\nu$. It must be noted that this formulation is arising in fluid dynamics, where in the original formulation $\nu$ is a vector field in $X$ and $\displaystyle{\int}\mid\mid \nu_{t}\mid\mid_{L^{2}(\mu_{t})}$ is the total kinetic energy. Here on the space of Borel probability measures on $X$. Now the natural task to carry out for our applications is to discretize the Wasserstein space and obtain a discretized version of the  Benamou-Brenier flow. To do this one can assume a triangulation of the space of probability distributions on $X$, that is $dens(X)$, is given. In Appendix. C. We have provided a detailed construction of the graph $G(x_{i}, a_{ij}), i,j=0,\cdots, m$ associated to $dens(X)$ on which we discretize the Wasserstein distance function. Assume the graph $G(x_{i}, a_{ij}), i,j=0,\cdots, m$ is given. We can associate a discretized "\textit{tangent sheaf}" to the graph, by assigning vector fields $\mu_{i}, \nu_{i}$ to the vertices,  and $\mu_{i}$ $\nu_{i,j}, i,j=1,n$ characterizing the amount transportation from node $i$ to $j$ over the edges. Using this construction the Wasserstein distance between two probability measures $\mu, \nu$ distributed over the graph is given as $$\langle\mu,\nu\rangle=\sum_{ij}W_{ij}(\mu_{i}, \nu_{j})=\sum_{i,j}c_{ij}(\rho)\mu_{i,j}\nu_{ij},$$where $c_{ij}(\rho)$ is the ``\textit{conductance coefficient}" associated to the density distribution $\rho$ on edge $e_{ij}$ with length $a_{ij}$.  Similarly, one defines the discretized inner product one can define the discretized divergence of a probability vector field distribution over the graph as
\begin{equation}
\text{div}_{\rho}(\nu):= \sum_{i\neq j}c_{ij}(\rho)\nu_{ji}
\end{equation}
Now the dynamic equation induced by Benamou-Brenier formalism is obtained as
\begin{equation}
\frac{d \mu_{t}}{dt}=\text{div}_{\rho^{t}}(\mu_{t}).
\end{equation}

\section{RG flow, and its connection to Wasserstein Gradient flow and diffusion dynamics}\label{RG-MK}
As described in Appendix. A. the exact renormalization group is the non-perturbative renormalization scheme that builds on optimal transport towards an uncorrelated Gaussian target distribution. The partition functional here is the sum of free energy action and the interaction action which can be written as
\begin{equation}
Z=\int \exp\left(-\frac{1}{2}x\cdot \Sigma^{-1}.x-S_{int}(x)\right)
\end{equation}
where $\Sigma$ is the covariance matrix between random variables $x$. The action $S(x)=S_{free}(x)+S_{int}(x)$ specifies a probability distribution of random variable $x$, that is $$\mu(x)=\frac{1}{Z}e^{-S(x)},$$with $Z=\int e^{-S(x)} dx$ the partition function defined above. Now under RG flow we obtain the differential equation \cite{Ma2020C2009.11880, Matsumoto2020R2011.14687, Cotler2023R2202.11737}
\begin{equation}\label{relation}
\partial_{t}\mu(x)=-\frac{\partial_{x}\cdot \partial_{t}\Sigma\cdot \partial_{x}}{2}\mu(x)-\partial_{x}\cdot \partial_{t}\Sigma\cdot \left( (\Sigma^{-1}\cdot x)\mu(x)\right)
\end{equation}
It can be seen that Equation \eqref{relation} leads to the diffusion dynamics equation induced by the Wasserstein gradient of the free energy functional we defined earlier. The trick is to define the metric used in the Fokker-Plank equation as 
\begin{equation}
g^{-1}=\frac{\partial_{t}\Sigma}{2}
\end{equation}
which is a positive definite matrix. Using this definition the exact RG flow equation \eqref{relation} can be re-written in terms of metric $g$ as
\begin{equation}\label{replaced}
\partial_{t}\mu(x)=\partial_{x}\cdot g^{-1}\cdot \partial_{x}\mu(x)+2\partial_{x}\cdot g^{-1}\cdot \left(\mu(x)\partial_{x}S_{free}(x)\right)
\end{equation}
The RG scheme must be such a design so that $$\partial_{t}\Sigma\leq 0$$ otherwise the metric $g$ will not be positive semi-definite, and the RG flow might be unstable. This means that the covariance of random variable $x$ must generally decrease under $RG$ flow, which is compatible with how we defined it in earlier sections of the current article. Comparing Equation \eqref{replaced} to Equation \eqref{Fokker}, we see that the corresponding functional for RG flow, must be given as 
\begin{equation}\label{RG-FP}
F[\mu]=\int \mu(x)(2 S_{free}(x)+\log \mu(x)) dx
\end{equation}
which up to a constant corresponds to the KL-divergence of the actual probability $\mu(x)$ with respect to the hypothetical probability distribution $q(x)$ given by $$q(x)=-\frac{e^{-2S_{free}(x)}}{2}.$$
The Equation \eqref{RG-FP} then clearly shows that the RG flow can be interpreted as the optimal transport gradient flow that minimizes the RG functional $F[\mu]$ along the Wasserstein geodesic of the probability distributions supported on the metric space equipped with the metric $g=-2(\partial_{t}\Sigma)^{-1}$. 

\section{Diffusion-based model}
\subsection{Forward Diffusion (Renormalization)}
We use construction in Section \ref{B-B}. That is we realize the diffusion equation as the gradient flow associated with the Wasserstein distance function induced by Benamou-Brenier equations. The induced diffusion-based model is the generative model inspired by non-equilibrium thermal dynamics. A diffusion model gradually introduces random noise into data using a sequence of Markov chain steps and is then trained to reverse the process for image generation.

Let $x_{0}$ be the data and $x_{1:T}$ be a sequence of noisy samples. The forward diffusion is defined \cite{ho2020denoising} by a Markov process
\begin{align}
q(x_{t}\mid x_{{t-1}})=\mathcal{N}(x_{t} \text{; } \sqrt{1-\beta_{t}}x_{t-1}, \beta_{t}\mathbb{I}),
\end{align}
where $\beta_{1:T}$ denote the noise variance schedule. Under reparameterization, with $\epsilon_{t\mid t-1} \sim \mathcal{N}(0, \mathbb{I})$ we obtain 
\begin{align}
    x_{t}=\sqrt{1-\beta_{t}}x_{t-1}+ \sqrt{\beta_{t}}\epsilon_{t\mid t-1}.\label{diff_img_forward}
\end{align}
The diffusion chain is generated auto-regressively such that
\begin{align}
q(x_{1:T}\mid x_{0})=\prod_{t=1}^{T} q(x_{t}\mid x_{t-1}).
\end{align}
In fact, marginalization to any time step is tractable
$$
q(x_{t}\mid x_{0})=\mathcal{N}(x_{t}\text{; } \sqrt{\bar{\alpha}_{t}}x_{0},(1-\bar{\alpha}_{t}) \mathbb{I}),$$
where $\alpha_{t}=1-\beta_{t}$ and $\bar{\alpha}_{t}=\prod_{s=1}^{t}\alpha_{s}$. 

\subsection{Reverse Diffusion Process (Generation)} When conditioned on $x_{0}$, the reverse diffusion is defined by the Markov process
\begin{align}
q(x_{t-1}\mid x_{t}, x_{0})=\mathcal{N}(x_{t-1}\text{; } \bar{\mu}(x_{t}, x_{0}), \bar{\beta}_{t} \mathbb{I}),
\end{align}
where we have
\begin{align}
    \bar{\mu}(x_{t}, x_{0})\!=\!\frac{\sqrt{\alpha_{t}}(1\!-\!\bar{\alpha}_{t\!-\!1})}{1\!-\!\bar{\alpha}_{t}}x_{t}\!+\!\frac{\sqrt{\bar{\alpha}_{t\!-\!1}}\beta_{t}}{1\!-\!\bar{\alpha}_{t}}x_{0}, \quad \bar{\beta}_{t}\!=\!\frac{1\!-\!\bar{\alpha}_{t\!-\!1}}{1\!-\!\bar{\alpha}_{t}}\beta_{t}.
\end{align}

The landmark study \cite{ho2020denoising} demonstrated the capability of diffusion models to generate high-quality images, sparking significant interest in this area. Studies pertinent to our work can be grouped into two main categories: (1) studies aiming to provide alternative diffusion (corruption) processes to the standard Gaussian process \cite{nachmani2021non,bansal2022cold, daras2023soft, chung2022come} and (2) studies that focus on enhancing the performance of diffusion models \cite{yang2023diffusion,zhang2022fast, song2020improved,song2021denoising, nichol2021improved}. In the next section, we propose the Frequency Domain Diffusion Model (FDDM), which lies at the intersection of the two, as we achieve performance improvements under FDDM. 


\section{Applications: Forward and reverse Diffusion in Fourier Space} \label{F-Image}
\subsection{Overview and Motivation}

In this section, we propose the Frequency Domain Diffusion Model (FDDM). FDDM is a novel diffusion-based approach for image generation that maps data between image and frequency domains using the discrete cosine transform (DCT) \cite{rao2014discrete}. Unlike the existing works that operate in the image domain, FDDM performs diffusion-based generative modeling in the frequency domain. This model is motivated by the fact that many natural images have a sparse representation in the frequency domain, which makes it easier to separate signals from noise. 

\begin{figure}[t]
    \centering
    \includegraphics[width=0.65\columnwidth]{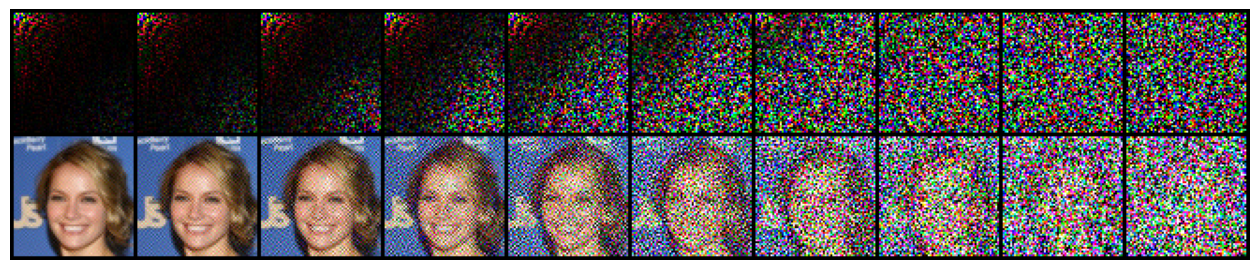}
    \caption{Scale-dependent noise schedule in the frequency domain (top) and the corresponding image domain representation, without patching, i.e., the whole image is a single patch. The frequency representation of the original (leftmost) image exhibits a sparsity, which we leverage in the proposed FDDM.}
    \label{fig:noise_schedule}
\end{figure}

During the forward process, we add noise to the image in the frequency domain, with the diffusion coefficient being determined by a novel scale-dependent noise schedule. We define this schedule using a dispersion relation, which describes the energy associated with each frequency component. The dispersion relation determines how quickly information diffuses through the image. The sparsity in the frequency domain means that most of the energy in an image is concentrated in a small number of frequency components, while the remaining components have very low energy (see Figure~\ref{fig:noise_schedule}). With this sparsity, we separate signal from noise more effectively than in image space. This is because the noise is spread out across all frequency components, while the signal is concentrated in a small number of components (unlike in the image domain). 

\begin{figure}[t]
    \centering
    \includegraphics[width=0.65\columnwidth]{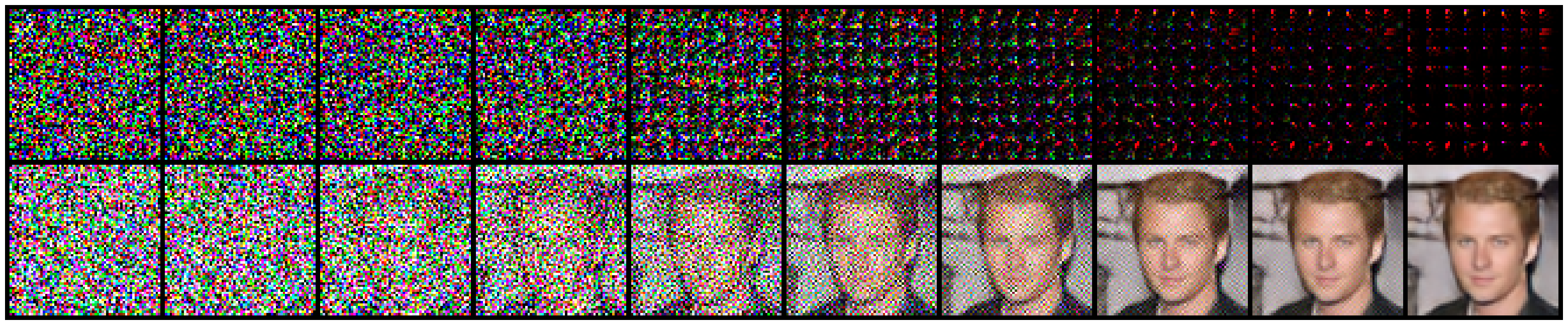}
    \caption{Image denoising and corresponding frequency domain representation for a $8\times8$ patched image. FDDM generates new samples by gradually denoising in the frequency domain.}
    \label{fig:sample}
\end{figure}

Another advantage of the FDDM is that it can handle both large-scale (low frequency) and small-scale (high frequency) features (components) in an image. By separating these in the frequency domain, we apply different diffusion coefficients to each scale. We then leverage ideas from JPEG encoding \cite{wallace1991jpeg, xu2020learning, gueguen2018faster} and apply FDDM on patches of images (see Figure~\ref{fig:sample}), significantly increasing the training/inference speed (compared to image domain diffusion), making FDDM suitable for time-critical applications such as medical imaging where there is a need for rapid image generation \cite{kazerouni2023diffusion}.  

In short, in this section
(1) We introduce a novel diffusion model, FDDM, that operates in the frequency domain using a scale-dependent noise schedule.
(2) We combine our process with ideas from JPEG encoding and frequency domain learning to refine the training protocol, thereby achieving improved speed in both training and inference, without a significant drop in generated image quality.

\subsection{Fourier-Space Diffusion Model}

We denote the data in the image space as $\vect{x}$ and in the frequency space as $\tilde{\vect{x}}$ such that $\tilde{\vect{x}}=\scF(\vect{x})\Leftrightarrow\vect{x}=\scF^{-1}(\tilde{\vect{x}}).$ 
The inverse frequency transformation $\scF^{-1}$ maps the data back to its original domain. In this work, we use the discrete cosine transformation (DCT) \cite{rao2014discrete} to ensure that real data maps to real features. DCT is similar to the Fourier transformation, using real-valued cosine functions instead of complex exponentials. It is more suitable for processing real-valued data like images.

Large-scale (small-scale) features correspond to low-frequency (high-frequency) components in the frequency domain. By leveraging this separation, we apply different diffusion coefficients to each scale to improve the denoising performance. In particular, we use momentum coordinates in the frequency domain and label the components of the transformed data by their momentum $\vect{k}$, defined as $\vect{k} = (k_1,k_2)$, where $k_1$ and $k_2$ are wave numbers that coordinate the frequency domain. Then we apply different diffusion coefficients based on the location in the frequency domain.

\subsubsection{Forward Diffusion in the Frequency Domain}

Unlike in the image domain, as in (\ref{diff_img_forward}), the forward diffusion process in FDDM gradually introduces noise from ultraviolet (UV), i.e., high frequency, to infrared (IR), i.e., low frequency, in the frequency domain (see Figure~\ref{fig:noise_schedule}), using a stochastic process $\tilde{\vect{x}}_0\to \tilde{\vect{x}}_1\to\cdots\to \tilde{\vect{x}}_t\to\cdots$ defined by 
$$\tilde{\vect{x}}_{t+1} = \sqrt{1 - \vect{\beta}_t} \odot \tilde{\vect{x}}_t + \sqrt{\vect{\beta}_t} \odot \vect{z}_t,$$
where we have $\vect{z}_t\sim\scN(\vect{0},I)$ and $\vect{\beta}_t$ is a scale-dependent diffusion coefficient that controls how much noise we introduce at each step. The symbol $\odot$ denotes element-wise multiplication. Alternatively, the same process can be defined using an $\bar{\vect{\alpha}}_t$ parameter that controls the SNR at each step:
\begin{align}
    \tilde{\vect{x}}_t = \sqrt{\bar{\vect{\alpha}}_t} \odot \tilde{\vect{x}}_0 + \sqrt{1 - \bar{\vect{\alpha}}_t} \odot \vect{z}_t,\label{eq: forward}
\end{align}
where we have $\vect{z}\sim\scN(\vect{0},I)$. We specify the design of $\bar{\vect{\alpha}}_t$ in Section~\ref{sec:noise_schedule}, where we discuss the proposed scale-dependent noise schedule. The key idea of this noise schedule is to gradually add correlated noise to the image from small-scale to large-scale in the forward process. By controlling the diffusion coefficient, i.e., SNR, based on locations in the frequency space, we essentially apply different amounts of smoothing to different scales and locations in the frequency domain.

\subsubsection{Backward Diffusion in the Frequency Domain}

The backward diffusion process in FDDM learns to generate data from IR to UV in the frequency domain, using a noise prediction model $\vect{\phi}_\theta$ that takes the noisy frequency features $\tilde{\vect{x}}_t$ and predicts the clean image in the frequency space $\tilde{\vect{x}}_0$ such that $\vect{\phi}_\theta(\tilde{\vect{x}}_t, t)\to \tilde{\vect{x}}_0$.
By utilizing the predicted clean image, we can perform the denoising operation. Following \cite{song2021denoising}, we define the following relationship between $\tilde{\vect{x}}_{t-1}$ and $\tilde{\vect{x}}_t$

\begin{align}
\tilde{\vect{x}}_{t-1} = \underbrace{ \sqrt{ \bar{\boldsymbol{\alpha}}_{t-1}} \odot \vect{\phi}_\theta(\tilde{\vect{x}}_t, t)  + \sqrt{ 1 - \bar{\boldsymbol{\alpha}}_{t-1} - \boldsymbol{\sigma}_{t}(\eta) } \odot \vect{\hat{z}} }_{\text{deterministic part}} + \underbrace{\boldsymbol{\sigma}_{t}(\eta) \odot \vect{z}_t}_{\text{stochastic part}},\label{eq:sampling}
\end{align}
with $\boldsymbol{\sigma}_{t}(\eta)\!=\!\eta \sqrt{\frac{1-\bar{\boldsymbol{\alpha}}_{t-1}}{1-\bar{\boldsymbol{\alpha}}_{t}}} \!\odot\! \sqrt{1\!-\!\frac{\bar{\boldsymbol{\alpha}}_{t}}{\bar{\boldsymbol{\alpha}}_{t-1}}}$. The predicted noise is
\begin{align}
    \hat{\vect{z}} \!=\! \dfrac{\tilde{\vect{x}}_{t} \!-\! \vect{\phi}_\theta(\tilde{\vect{x}}_t, t)}{\sqrt{ 1 \!-\! \bar{\boldsymbol{\alpha}}_{t}}}.
\end{align}
The backward diffusion process learns to generate cleaner data from noisy data in the frequency space by predicting the noise configuration and using it to recover the clean signal (see Figure~\ref{fig:sample}). As in the forward process, in the backward process, we apply different amounts of smoothing to different scales and locations in the frequency space by controlling the diffusion coefficient (or SNR) based on momentum coordinates.

\subsubsection{Objective Function and Training Approach}

The objective function for training the noise prediction model $\vect{\phi}_\theta$ in FDDM is
\begin{align}
    \mathcal{L}_\theta = \mathop{\dsE}_{{\vect{x}}_0 \in \mathcal{D}} \mathop{\dsE}_{\vect{z}\sim\scN(\vect{0},I)} \mathop{\dsE}_{t \sim \mathcal{U}(1,\ldots, T)} \|(\tilde{\vect{x}}_0 - \vect{\phi}_\theta(\tilde{\vect{x}}_t, t))\|^2, \label{eq:loss}
\end{align}
where $\mathcal{D}$ is the set of training images. ${\vect{x}}_0$ is the image drawn from the training set, and $\tilde{\vect{x}}_t$ is the noisy frequency data obtained through the forward diffusion described by (\ref{eq: forward}) given a noise configuration $\vect{z}$. The objective function measures the mean squared error between the predicted clean image $\vect{\phi}_\theta(\tilde{\vect{x}}_t, t)$ and the true clean image $\tilde{\vect{x}}_0$ over all training images. We note that the loss in (\ref{eq:loss}) is computed in the frequency domain. Similarly, in FDDM, the forward and backward processes take place in the frequency domain. 

We use Adam optimizer \cite{kingma2014adam} to minimize the loss function in (\ref{eq:loss}) by sampling a mini-batch of training images from $\mathcal{D}$ and computing the gradients with respect to the parameters of the noise prediction model. Adam updates the parameters using a learning rate and adjusts the rate for each parameter based on the first and second moments of the gradients, improving the optimization's convergence properties. By updating the parameters iteratively over many epochs, we learn a noise prediction model that denoises images in the frequency domain.

\subsubsection{Proposed Scale-Dependent Noise Schedule}\label{sec:noise_schedule} 
The scale-dependent noise schedule in FDDM is a momentum-dependent function that controls the amount of noise introduced at each step of the forward process. We define it using a dispersion relation $\epsilon_\vect{k}$, which gives the energy associated with a frequency mode of momentum $\vect{k}$. We use a tight-binding dispersion such that $\epsilon_\vect{k} =  - \cos{\pi k_1} - \cos{\pi k_2}.$ In particular, the scale-dependent noise schedule is given by
\begin{align}
    {\bar{\alpha}}_{t,\vect{k}} = \frac{1}{\exp\left(\frac{\epsilon_\vect{k} - \mu_t}{T'}\right) + 1},
\end{align}
where $\mu_t$ is the Fermi level at time $t$, which controls the overall SNR. The noise schedule varies with momentum $\vect{k}$, and can be packed into a vector $\bar{\vect{\alpha}}_t = [{\bar{\alpha}}_{t,\vect{k}}]_{\vect{k} \in \textsf{BZ}}$, where $\textsf{BZ}$ denotes the Brillouin zone. The Fermi level $\mu_t$ is expected to decrease with diffusion time $t$, such that $\bar{\vect{\alpha}}_t$ decreases from 1 to 0 as more noise is introduced. This allows for a gradual introduction of noise from UV to IR in the frequency space. The temperature parameter $T'$ controls how sharp or smooth this transition is, with lower values of $T'$ leading to sharper transitions. With this scale-dependent noise schedule, we essentially apply different amounts of smoothing to different scales and locations in the frequency space. This allows for effective denoising while preserving important features in images.

\subsection{Description of FDDM}\label{sec:implementation}
FDDM consists of two algorithms: frequency-based forward process and sampling, which are given in Algorithms~\ref{algo:train} and~\ref{algo:samp}, respectively. Frequency-based forward process includes pre-processing, forward process, and training steps. 

\begin{figure*}[t]
\centering
\includegraphics[width=0.85\textwidth]{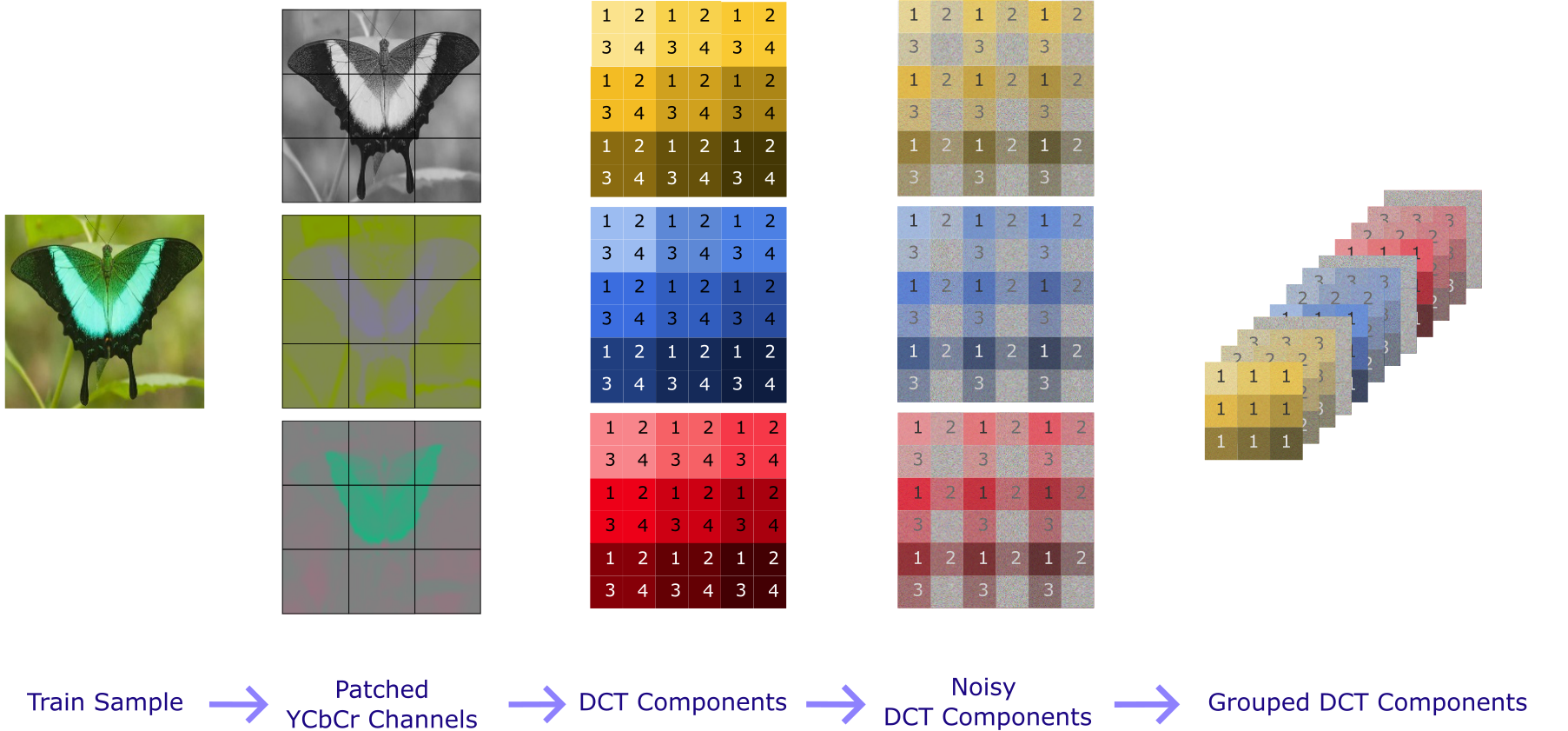}    
\caption{The proposed frequency-based forward process. We split YCbCr input image channels into patches and take the DCT of each patch. Then, we add the scale-dependent noise to each patch in the frequency domain. Finally, we group the DCT components into output channels, with each channel containing components of the same frequency. We feed these output channels, i.e., grouped DCT components to the diffusion model during training.}\label{fig:forward_process}
\end{figure*}

\noindent \textbf{Frequency-Based Forward Process.} We first sample an original image from a set $\mathcal{D}$ of training images. This sample image is initially in the RGB format. Following the standard practice of JPEG encoding \cite{xu2020learning}, we first convert it to YCbCr format and split it into patches to obtain enhanced training and inference speed. We note that this patch-based forward process is enabled by the FDDM, as it performs diffusion in the frequency domain. Next, we take the DCT of each patch separately. We then sample a timestep uniformly for the image and corrupt this image based on the forward process defined in (\ref{eq: forward}). In particular, we apply the scale-dependent noise to each patch independently. Once the noise is applied, we obtain the noisy DCT components for each patch. The final step of the pre-processing is to group the components from the same frequency across patches into separate channels. In our design, this process is performed for a batch of images and we feed the associated uniformly sampled timestep of each image of the batch, i.e., noise level, to the neural network, together with the corrupted (and grouped) DCT components, i.e., channels. Finally, we take a gradient descent step using the L2 loss between the actual input image in the frequency domain and the predicted clean image in the frequency domain. This entire patchifying and forward process pipeline is demonstrated in Figure~\ref{fig:forward_process} and Algorithm~\ref{algo:train} (which does not explicitly state the grouping of the noisy DCT components for ease of exposition). 

We let $n$ denote the size of the images, i.e., input images are of size $n\times n$. Assume the patching operation is performed with a patch size of $d \times d$. Then, the resulting patchified image has a total of $P=\left(\frac{n}{d}\right)^2$ patches. We note that this operation is repeated for each input channel (YCbCr images have 3 channels). In the example in Figure~\ref{fig:forward_process}, the image size is $n=6$ and the patch size is $d=2$. Thus, we have a total of $P=9$ patches in each image channel and the resulting tensor is of size $12\times 3 \times 3$, where $12$ is the number of output channels and $3 \times 3$ is the size of each output channel. We note that in the standard JPEG convention \cite{wallace1991jpeg,xu2020learning}, patches are of size $8 \times 8$. Also, we note that the patch size directly determines the number of DCT components we have in each patch which is $d^2=4$, as also demonstrated in Figure~\ref{fig:forward_process}.

\begin{algorithm}[h]
    \caption{Frequency-Based Forward Process} \label{algo:train}
    \begin{algorithmic}[1]
    \Require Input distribution $D$, \# of training timesteps $T$, patch size $d$, image size $n$
    \State $P \gets (n / d)^{2} $ 
    \Comment{Total number of patches}
    \Repeat
        \State $\vect{x}_0 \sim \mathcal{D}$
        \State $t \sim \mathcal{U}(1,\ldots, T)$
        \State Split $\vect{x}_0$ into $d\times d$ patches $\{\vect{x}_0^p \}_{p \in \{1,\ldots, P\}}$
        \For{$p \in \{1,\ldots, P\}$}
            \State $\vect{z} \gets \mathcal{N}(\vect{0}, \vect{I_d})$
            \State $\tilde{\vect{x}}_0^p \gets \scF(\vect{x}_0^p)$
            \State $\tilde{\vect{x}}_t^p \gets \sqrt{\bar{\vect{\alpha}}_t} \odot \tilde{\vect{x}}_0^p + \sqrt{1 - \bar{\vect{\alpha}}_t} \odot \vect{z}$
        \EndFor
        \State $\tilde{\vect{x}}_0 \gets [\tilde{\vect{x}}_0^0, \tilde{\vect{x}}_0^1, \dots, \tilde{\vect{x}}_0^P  ]$ \Comment{Concatenate patches}
        \State $\tilde{\vect{x}}_t \gets [\tilde{\vect{x}}_t^0, \tilde{\vect{x}}_t^1, \dots, \tilde{\vect{x}}_t^P  ]$
        \State Take gradient descent step on $\nabla_{\theta}\|\tilde{\vect{x}}_0 - \vect{\phi}_\theta({\tilde{\vect{x}}}_t, t)\|^2$
    \Until{converged}
\end{algorithmic}
\end{algorithm}

\noindent\textbf{Sampling (Inference). }Sampling is essentially the backward process (denoising). We start by sampling pure noise in the frequency domain and gradually denoise it using the trained model, following (\ref{eq:sampling}). Algorithm~\ref{algo:samp} shows the denoising operation, which is also shown in Figure~\ref{fig:sample}. We note that, as in the forward process, we work on patches in the backward process. That is, we perform the denoising operation described by (\ref{eq:sampling}) on the grouped DCT components, which are omitted in Algorithm ~\ref{algo:samp} for ease of exposition. In particular, using the model prediction at timestep $t_i$, $\vect{\phi}_\theta(\tilde{\vect{x}}_{t_i}, t_i)$, we recover the less noisy DCT components from the previous timestep $t_{i-1}$, and these steps are repeated until timestep $t_1$. Once the denoising is completed, we take the inverse DCT of the patches, combine the results (unpatch the image), and return the generated image. Here, $t_i$ denotes the inference time steps for $i \in \{1, \ldots, I\}$, with $I$ being the total number of inference steps. Unlike the forward process, where the timesteps are consecutive, in the backward process, we can sparsely sample. That is, $t_i$ and $t_{i-1}$ are not necessarily consecutive for any $i \in \{1, \ldots, I\}$. Ideally, the resulting generated image should look as if it is from the original dataset. 

\begin{algorithm}[t]
    \caption{Sampling (Inference)} \label{algo:samp}
    \begin{algorithmic}[1]
        \Require Inference time steps $\{t_i\}_{i \in \{1, \dots, I\}}$,  total \# of inference steps $I$, patch size $d$, image size $n$, $\eta$
        \State $P \gets (n / d)^{2} $ 
        \State $\{\tilde{\vect{x}}_0^p \gets \mathcal{N}(\mathbf{0}, \vect{I_d}) \}_{p \in \{1,\ldots, P\}}$
        \State $\tilde{\vect{x}}_{t_{I}} \gets [\tilde{\vect{x}}_{t_{I}}^0,\tilde{\vect{x}}_{t_{I}}^1, \dots ,\tilde{\vect{x}}_{t_{I}}^P]$
        \For{$i = I$ \textbf{downto} $1$}
            \State $[\tilde{\vect{x}}_{0}^0,\tilde{\vect{x}}_{0}^1, \dots ,\tilde{\vect{x}}_{0}^P] \gets \vect{\phi}_\theta({\tilde{\vect{x}}}_{t_i}, t_i)$ 
            
            
            \If {$i > 1$}
            \For{$p \in \{1,\ldots, P\}$}
                \State $\vect{z} \gets \mathcal{N}(\vect{0}, \vect{I_d})$
                \State $ \boldsymbol{\sigma}_{t_i}(\eta)=\eta \sqrt{\dfrac{1-\bar{\boldsymbol{\alpha}}_{t_{i-1}}}{1-\bar{\boldsymbol{\alpha}}_{t_i}}} \odot \sqrt{1-\dfrac{\bar{\boldsymbol{\alpha}}_{t_i}}{\bar{\boldsymbol{\alpha}}_{t_{i-1}}}}$
                \State $\hat{\vect{z}} = \dfrac{\tilde{\vect{x}}_{t_i}^p - \tilde{\vect{x}}_0^p}{\sqrt{ 1 - \bar{\boldsymbol{\alpha}}_{t}}}    $ 
                
                
                \State ${\vect{x}}_{t_{i-1}}^p =  \sqrt{ \bar{\boldsymbol{\alpha}}_{t_{i-1}}} \odot \tilde{\vect{x}}_{0}^p + \boldsymbol{\sigma}_{t_{i}}(\eta) \odot \vect{z}$
                \Statex $\hspace{8em}  +\sqrt{ 1 - \bar{\boldsymbol{\alpha}}_{t_{i-1}} - \boldsymbol{\sigma}_{t_{i}}(\eta) } \odot \vect{\hat{z}}$
            \EndFor
            \EndIf
            \State $\tilde{\vect{x}}_{t_{i-1}} \gets [\tilde{\vect{x}}_{t_{i-1}}^0,\tilde{\vect{x}}_{t_{i-1}}^1, \dots ,\tilde{\vect{x}}_{t_{i-1}}^P]$
        \EndFor
        \State $\{ \vect{x}_0^p \gets \scF^{-1}(\tilde{\vect{x}}_0^p)\}_{p \in \{1,\dots,P\}}$
        \State \Return combined (unpatched) $[\vect{x}_{0}^0,\vect{x}_{0}^1, \dots ,\vect{x}_{0}^P]$
    \end{algorithmic}
\end{algorithm}

\subsection{Experiment Results}

The Fermi level $\mu_t$ is linearly decreasing over time $t$ from $4$ to $-4$. The momentum coordinates in the frequency domain $\vect{k} = (k_1,k_2)$ take values between $0$ and $1$ with linear steps, where the step size is determined by the input image size. For example, for an image from the CelebA-64 \cite{celebA} dataset of size $64 \times 64$, the precision of momentum coordinate increments is $\frac{1}{64}$. By using these momentum coordinates, we effectively vary the applied noise intensity on the input image. We set the temperature parameter $T'$ to $0.5$. 
\begin{figure}[t]
    \centering
        \includegraphics[width=0.35\columnwidth]{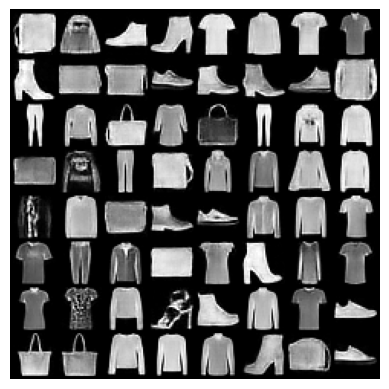}
        \caption{Uncurated generated samples on Fashion-MNIST.}
\label{fig:uncurated_fashion}
        \vspace{-5mm}
\end{figure}

\begin{figure*}[t]
    \centering
    \begin{minipage}{0.49\textwidth}
        \centering
        \includegraphics[width=0.86\textwidth]{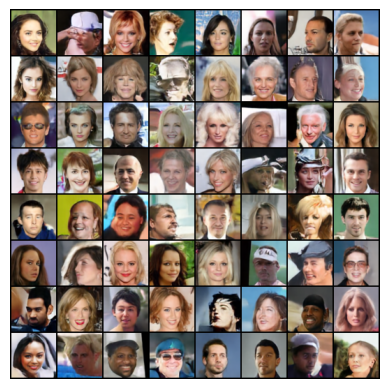}
        \caption{Uncurated samples on CelebA-64 dataset.}
        \label{fig:uncurated}
    \end{minipage}%
    \begin{minipage}{0.49\textwidth}
        \centering
        \includegraphics[width=0.86\textwidth]{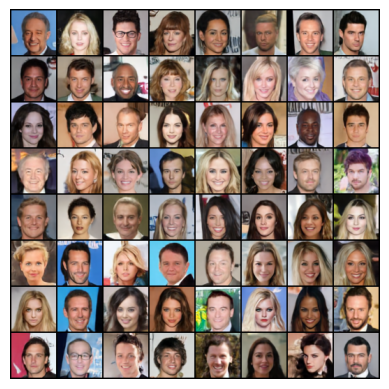}
        \caption{Curated samples on CelebA-64 dataset.}
        \label{fig:curated}
    \end{minipage}%
    \vspace{-5mm}
\end{figure*}

In their seminal diffusion work in \cite{ho2020denoising}, authors utilize a U-Net structure based on ResNet blocks with downsampling followed by an upsampling process and as well as an attention module in between the convolutional blocks. 
Inspired by \cite{xu2020learning}, we modify the U-Net structure in \cite{ho2020denoising} to match the output of our frequency-based forward process. First, we change the number of input channels from 3 (RGB channels) to $3 \times d \times d$ (number of frequency channels). Second, we keep the resolution constant throughout the model. Since the neural network parameters are shared across time, we use sinusoidal position embedding to encode time so that the neural network knows at which noise level it operates during the denoising process. In addition, since in FDDM the noise is scale-dependent, by inputting the time, we implicitly feed the noise-scale to the neural network as well. 

To report the performance of the proposed FDDM, we perform experiments on Fashion-MNIST \cite{fashion_mnist} (resolution $28 \times 28)$ and  CelebA-64 (resolution $64\times64$) \cite{celebA}. We configure the forward and reverse process using $\eta=1$ and $T=I=1000$. We use a batch size of 128 and train for a total of 800k steps. We use the Adam optimizer in training, with the learning rate set to $2\times10^{-4}$, without any sweeping. We carry out the experiments using NVIDIA A100 40GB. 

We first present a set of uncurated samples generated by FDDM, using Fashion-MNIST with patch size $d=7$ in Figure~\ref{fig:uncurated_fashion}, demonstrating its image generation capabilities in the frequency domain. Next, we consider a more realistic CelebA-64 dataset and present the curated and uncurated generated samples in Figs.~\ref{fig:uncurated} and~\ref{fig:curated} for $d=4$. Steps of the image denoising process and the corresponding frequency domain representation are as in Figure~\ref{fig:sample}, showcasing denoising in UV and IR parts as a function of timesteps. Overall, these results for both datasets demonstrate that the proposed FDDM successfully generates compelling images.

Next, we compare the performance of FDDM with the seminal image domain diffusion approach of \cite{ho2020denoising} named DDPM, as the FDDM architecture is based on DDPM and our goal is to offer a performance-utility trade-off for this design. For comparisons, we use MACs (Multiply-accumulate operations) \cite{yang2023diffusion}, number of inference steps per second, and Fréchet Inception Distance (FID) \cite{NIPS2017_8a1d6947} as our metrics. FID measures the similarity between two sets of images and is shown the correlate with human judgement. In general, a low FID score indicates a good generated sample quality. 

First, we take a look at the number of inference steps per second and MACs of the two schemes. Results in Table~\ref{table:comparison} show that inference in our proposed FDDM is significantly faster than DDPM (around 2.7 to 8.5$\times$ faster). This is attributed to our method of patching and forming frequency-based channels (as explained in section~\ref{sec:implementation}), which effectively reduces the computations in the U-Net. The performance enhancement is evident from the MACs column (lower the better) in Table~\ref{table:comparison}, demonstrating that our model requires significantly fewer MACs to generate 128 samples.

We compute the FID scores for the CelebA-64 dataset using 200000 samples and show the results in Table~\ref{table:comparison}. As expected, our FID scores are higher than the baseline, considering our faster and more time-efficient design. These results indicate a possible trade-off between the sample quality and training runtime efficiency. One observation is that lower FID scores around 3 are achieved when the finer details of the images are present. This means that our FID score around 18 does not mean $6\times$ worse images, and in fact the images generated by the proposed FDDM may be suitable for certain downstream tasks, for which finer details are not as critical. For example, the images we generate in Figure~\ref{fig:curated} can be part of a synthetic dataset for training a classifier model that groups people according to their certain physical characteristics, e.g., hair color, glasses vs no glasses, and so on). For this task, finer details such as the background objects may not be as critical. 

\begin{table}[ht]
\centering
\begin{tabular}{lcccc}
\hline
Model &  MACs (G)  $\downarrow$ & \# inference steps/s  $\uparrow$ & FID score  $\downarrow$ \\
\hline
DDPM \cite{ho2020denoising} & 3099 & 5.344 & 3.26 \\
Ours (4x4 patch) & 1781 & 14.655 & 18.17\\
Ours (8x8 patch) & 449 & 45.76 & 20.65 \\
\hline
\end{tabular}
\caption{Performance comparison of the proposed FDDM with DDPM \cite{ho2020denoising} for $T=I=1000$ on CelebA-64. }
\label{table:comparison}
\end{table}

\appendix

\section{Discretization of Wasserstein distance function}
We would now elaborate on construction of a graph $G(x_{i}, a_{ij}), i,j=0,\cdots, m$ associated to $dens(X)$ which induces the discretized Wasserstein distance function, this will lead to derivation of a discretized Benamou-Brenier dynamics equation on space of probability distributions on $X$.

\begin{figure}[htbp]
\begin{center}
\includegraphics[width=0.9\columnwidth]{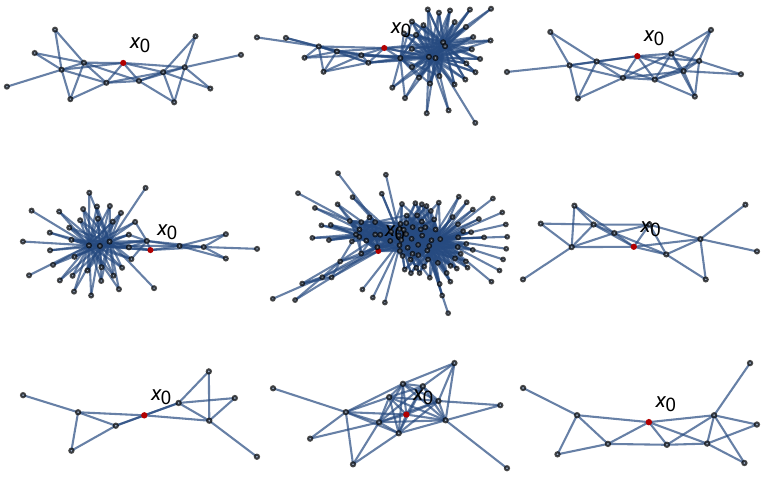}
\caption{Samples of the local graph around a randomly picked point $x_0$.}
\label{fig:graphs}
\end{center}
\end{figure}

 We would first define the notion of Ricci curvature associated to a discrete graph induced by a lattice. Consider the discrete lattice $\mathbb{Z}^{n}$. Fix a point $x_{0}\in \mathbb{Z}^{n}$ with coordinates $x:=(t_{1}, \cdots, t_{n}), t_{i}\in \mathbb{Z}$.  Now choose a number $\epsilon$ and consider $n$-dimensional balls $B_{x_{0}}(\epsilon)$ defined as set of points $x \in\mathbb{Z}^{n}$ such that $d(x,x_{0})<1+\epsilon$ where $d(x,x_{0})$ is the distance function with respect to the Eculidean metric in $\mathbb{Z}^{n}$. The latter inequality picks out a set of points in the lattice in $\epsilon$-neighborhood of the point $x_{0}$. We connect them to $x_{0}$ and label them as $x_{i}, i=1,2,\cdots$, assuming that $x_{1}$ satisfies closest Euclidean distance to $x_{0}$. Now we repeat the same procedure for points $x_{i}, i=1,\cdots$. Let us assume that, via iterating this procedure we obtain a graph $G(x_{i}, a_{ij}), i,j=0,\cdots, m$ with the central point $x_{0}$ as shown in Figure \ref{fig:graphs}. 

Now pick a vertex $x_{i}$ in the graph and an edge with length $a_{ij}$ connecting $x_{i}$ and $x_{j}$ for some $j\neq i$. We define the Wasserstein transport distance function induced by defining an optimal transport problem associated to this graph. We decorate the vertices of the graph with optimal transport weights as follows: We decorate $x_{i}, x_{j}$ with $\epsilon, 1-\epsilon$. In other words we associate to the vertex $x_{i}$ a probability distribution $\Psi_{x_{i}}(x,\epsilon)$ with most of its weight at $x=x_{i}$ and a small amount of weight at neighboring vertices, so that the average distance from $x_{i}$ to a vertex chosen from this distribution is much less than an edge length. Here the probability distribution function is defined by

\[   
\Psi_{x_{i}}(x,\epsilon) = 
     \begin{cases}
       1-\frac{d_{J}(x_{i})}{D_{x_{i}}}\epsilon &\quad\text{if $x=x_{i}$}\\
       \frac{J_{x_{i}x}}{D_{x_{i}}}\epsilon &\quad\text{if $x\simeq x_{i}$}\\
       0&\quad\text{Otherwise},\\
       \end{cases}
\]
where $d_{J}(x_{i})=\sum_{x\simeq x_{i}}J_{x_{i}x}$, $D_{x_{i}}$ is the lapse function which determines how fast $\epsilon$ runs in different locations of the graph, and $J_{xy}=\frac{1}{a^{2}_{xy}}$. Now simplifying this for our situation, for $x_{k}, k\neq i,j$ we assign probability weight functions $\epsilon_{k|_{i}}:= \Psi_{x_{i}}(x_{k},\epsilon)$ which are given by$$\displaystyle{\epsilon_{k|_{i}}:=\epsilon\cdot\frac{\bigg(\frac{1}{a^{2}_{ki}}\bigg)}{\sum_{m\neq i}\frac{1}{a^{2}_{im}}}},$$where $a_{ki}$ are defined as the edge lengths obtained in the previous step. Now we define the Wasserstein transport distance function associated to the edge $a_{ij}$
\begin{equation}\label{transport}
W_{ij}=(1-\epsilon-\epsilon_{j})+\sum_{k\neq j,i}\epsilon_{k}\frac{(a_{ik}+a_{ij})}{a_{ij}}
\end{equation}

Now the Ricci curvature of the edge $d_{ij}$ is defined by 
\begin{equation}\label{limit}
k_{ij}:=\lim_{\epsilon\to 0}\frac{1-W_{ij}}{\epsilon}.
\end{equation}
Next one can calculate the Ricci curvature of the manifold passing through point $x_{i}$ in  ball of radius $1+\epsilon$ centered at $x_{0}$. The local curvature of the manifold is defined as the product of the maximum and minimum values attained by $k_{ij}$ for all $x_j$ adjacent to the vertex $x_{i}$. 

Once the curvature is obtained the manifold can be constructed by gluing local patches with the given curvature. 

\textbf{Example:} Let us assume that the obtained graph is as given in figure below. 

\begin{center}
\tikzset{every picture/.style={line width=0.80pt}} 

\begin{tikzpicture}[x=0.75pt,y=0.75pt,yscale=-1,xscale=.6]

\draw    (85,81) -- (281,180) ;

\draw    (281,180) -- (422,98) ;

\draw    (435,257) -- (281,180) ;

\draw    (168,275) -- (281,180) ;

\draw (285,196) node   {$x_{0}$};
\draw (282,180) node   {$\bullet$};

\draw (424,85) node   {$x_{1}$};
\draw (431,267) node   {$x_{2}$};
\draw (165,281) node   {$x_{3}$};
\draw (170,275) node   {$\bullet$};

\draw (78,91) node   {$x_{4}$};
\draw (85,81) node   {$\bullet$};

\draw (331,129) node   {$a_{0,1}$};
\draw (392,217) node   {$a_{0,2}$};
\draw (212,214) node   {$a_{0,3}$};
\draw (205,120) node   {$a_{0,4}$};
\draw (450,100) node   {$\epsilon _{1|_{0}}$};
\draw (420,100) node   {$\bullet$};

\draw (464,267) node   {$\epsilon _{2|_{0}}$};
\draw (434,257) node   {$\bullet$};

\draw (128,275) node   {$\epsilon _{3|_{0}}$};
\draw (110,78) node   {$\epsilon _{4|_{0}}$};

\end{tikzpicture}
\end{center}

Let us compute the curvature of $E_{0,1}$ which connects $x_{0}$ and $x_{1}$. We associate to $x_{0}, x_{1}$ the probability weight function $\epsilon, 1-\epsilon$ respectivtely. Now use the identity above the values of $\epsilon_{2}, \epsilon_{3}, \epsilon_{4}$ are given as

\begin{align}
&
\epsilon_{0}=1-\epsilon\notag\\
&
\epsilon_{1|_{0}}=\epsilon\notag\\
&
\epsilon_{2|_{0}}=\epsilon\cdot\frac{\frac{1}{a^{2}_{0,2}}}{\frac{1}{a^{2}_{0,2}}+\frac{1}{a^{2}_{0,3}}+\frac{1}{a^{2}_{0,4}}}\notag\\
&
\epsilon_{3|_{0}}=\epsilon\cdot\frac{\frac{1}{a^{2}_{0,3}}}{\frac{1}{a^{2}_{0,2}}+\frac{1}{a^{2}_{0,3}}+\frac{1}{a^{2}_{0,4}}}\notag\\
&
\epsilon_{4|_{0}}=\epsilon\cdot\frac{\frac{1}{a^{2}_{0,4}}}{\frac{1}{a^{2}_{0,2}}+\frac{1}{a^{2}_{0,3}}+\frac{1}{a^{2}_{0,4}}}
\end{align} 
Now using \eqref{transport}, we compute the Wasserstein transport distance function associated to the edge $E_{01}$ is given by 

\begin{align}
W_{01}=(1-\epsilon-\epsilon_{1|_{0}})+\epsilon_{2|_{0}}\frac{(a_{0,2}+a_{0,1})}{a_{0,1}}+\epsilon_{3|_{0}}\frac{(a_{0,3}+a_{0,1})}{a_{0,1}}+\epsilon_{4|_{0}}\frac{(a_{0,4}+a_{0,1})}{a_{0,1}}
\end{align}
Which can then induce the curvature of $E_{0,1}$ using \eqref{limit}. When applying our formalism to Benamou-Brenier formalism we obtain the following. Consider the metric graph $G(x_{i}, a_{ij}), i,j=0,\cdots, m$ constructed over $dens(X)$ with probability measures $\mu$ and $\nu$.

\bibliographystyle{plain}
\bibliography{ref}

\noindent{\small{\tt{artan@mit.edu, yzyou@physics.ucsd.edu, buyukate@usc.edu, 
ziashaha@usc.edu, avestime@usc.edu}}

\end{document}